\documentclass[prb,twocolumn,floatfix,groupedaddress]{revtex4}

\usepackage{graphicx,amsfonts,amssymb,hyperref,nicefrac,lipsum, verbatim}

\usepackage[fleqn]{amsmath}
\usepackage{verbatim}
\usepackage{mathtools}

\usepackage[x11names]{xcolor}
\newif\ifhyper
\hypertrue
\ifhyper
\hypersetup{
   citecolor = {green!60!black},
   colorlinks = {true}, % false
   urlcolor = {blue} % magenta
}
\fi

\newcommand{\be}{\begin{equation}}
\newcommand{\ee}{\end{equation}}
\newcommand{\beqa}{\begin{eqnarray}}
\newcommand{\eeqa}{\end{eqnarray}}

\def\Longarrow{\protect\@lra}
\def\@lra{\relbar\joinrel\relbar\joinrel\relbar\joinrel%
          \relbar\joinrel\rightarrow}

\begin{document}

\title{Delocalization of edge states in topological phases}

%\author{M.~Malki \and G.\ S.~Uhrig}
%\institute{                    
%  Lehrstuhl f\"ur Theoretische Physik 1, TU Dortmund, Germany
%}
%\pacs{73.20.Jc}{Delocalization processes}
%\pacs{03.65.Vf}{Phases: geometric; dynamic or topological}
%\pacs{71.23.An}{Theories and models; localized states}

\author{M.~Malki}
\email{maik.malki@tu-dortmund.de}
\affiliation{Lehrstuhl f\"ur Theoretische Physik 1, TU Dortmund, Germany}

\author{G.\ S.~Uhrig}
\affiliation{Lehrstuhl f\"ur Theoretische Physik 1, TU Dortmund, Germany}
\date{\rm\today}

\begin{abstract}
The presence of a topologically non-trivial discrete invariants implies the existence
of gapless modes in finite samples, but it does not necessarily imply their localization. 
The disappearance of the indirect energy gap in the bulk generically leads to the absence of 
localized edge states. 
We illustrate this behavior in two fundamental lattice models on the
single-particle level. By tuning a hopping parameter the indirect gap is closed while maintaining the topological properties. The inverse participation ratio is used to 
measure the degree of localization. 
\end{abstract}

\maketitle

Topological phases \cite{hasan10, qi11, burko11, recht13, goldm16} constitute
one of the most spectacular research fields in quantum matter. 
Historically, the earliest widely studied example is the quantum Hall effect
\cite{klitz80,storm83,niu85,hatsu93}. More recently,  topological insulators  have attracted
much interest \cite{hasan10, berne13}. The edge states in two dimensions (2D) \cite{hatsu93} 
and the surface states in three dimensions (3D) \cite{fu07}
in topological insulators are commonly seen as a characterizing feature.
For notational simplicity, we will henceforth use the term `edge state' for all states
localized at a boundary irrespective of dimensionality.
Such states have potential
 applications in spintronics \cite{wolf01}, magneto-electronics \cite{yue16} and opto-electronics \cite{yue17}. The application of the integer quantum Hall effect
in high-precision metrology stands out \cite{weis11}.
Another interesting suggestions are tunable group velocities of edge states to realize 
delay lines and interference devices \cite{uhrig16, malki17c}.

The emergence of edge states in non-interacting topological systems is
 elucidated by the bulk-boundary correspondence 
\cite{mong11, fidko11, fukui12, berne13}
which relates finite discrete topological invariants of the energy bands 
in the bulk to the existence of edge states at the boundary of finite systems. 
The underlying idea is as follows.
The transition between two bulk systems (one could be the vacuum) 
with different discrete topological invariants
cannot be continuous because of their discrete nature. Thus there must be
in-gap states which link the bands of different 
topological invariants so that they can no longer be defined for each band separately.
Since this argument hinges on the existence of the boundary, it is
assumed that these in-gap states are localized at the boundaries, hence
represent edge states \cite{berne13}. For certain Hamiltonians this
can be rigorously shown \cite{mong11, fidko11, fukui12}.

Such topological edge states can be found in topological insulators 
\cite{berne13,goldm13}, 
topological semi-metals \cite{wan11}, topological crystalline insulators \cite{fu11}.
Higher-order topological insulators in 3D may not display surface states, but so-called 
hinge states \cite{schin18}.
 In one dimension (1D), there can be localized states at the
chain ends \cite{kitae01, joshi17}. 
Recently, however, we found in 1D that localized end states do not represent
the generic scenario if the indirect energy gap between the bands of different
topological invariant vanishes \cite{malki18b}. While the direct gap $\Delta_\text{dir}$
measures the energetic separation of two bands at given fixed momentum, 
the indirect gap $\Delta_\text{indir}$ 
measures this separation  if momentum  changes are admitted.
Clearly, $\Delta_\text{indir}\le \Delta_\text{dir}$ and a finite $\Delta_\text{dir}$
is sufficient for the bands to be well-defined. This surprising finding
qualifies the bulk-boundary correspondence in the sense that 
a finite direct gap does not suffice to guarantee  localized edge states. 

Since 1D topological systems differ significantly from their higher
 dimensional counter parts, the question arises to which extent the delocalization 
of edge states occurs in 2D as well if the indirect gap vanishes.
The goal of the present Letter is to answer this question by a 
representative proof-of-principle study. 

%\section{Delocalization of edge states}

\begin{figure}
	\centering
		\includegraphics[width=1.0\columnwidth]{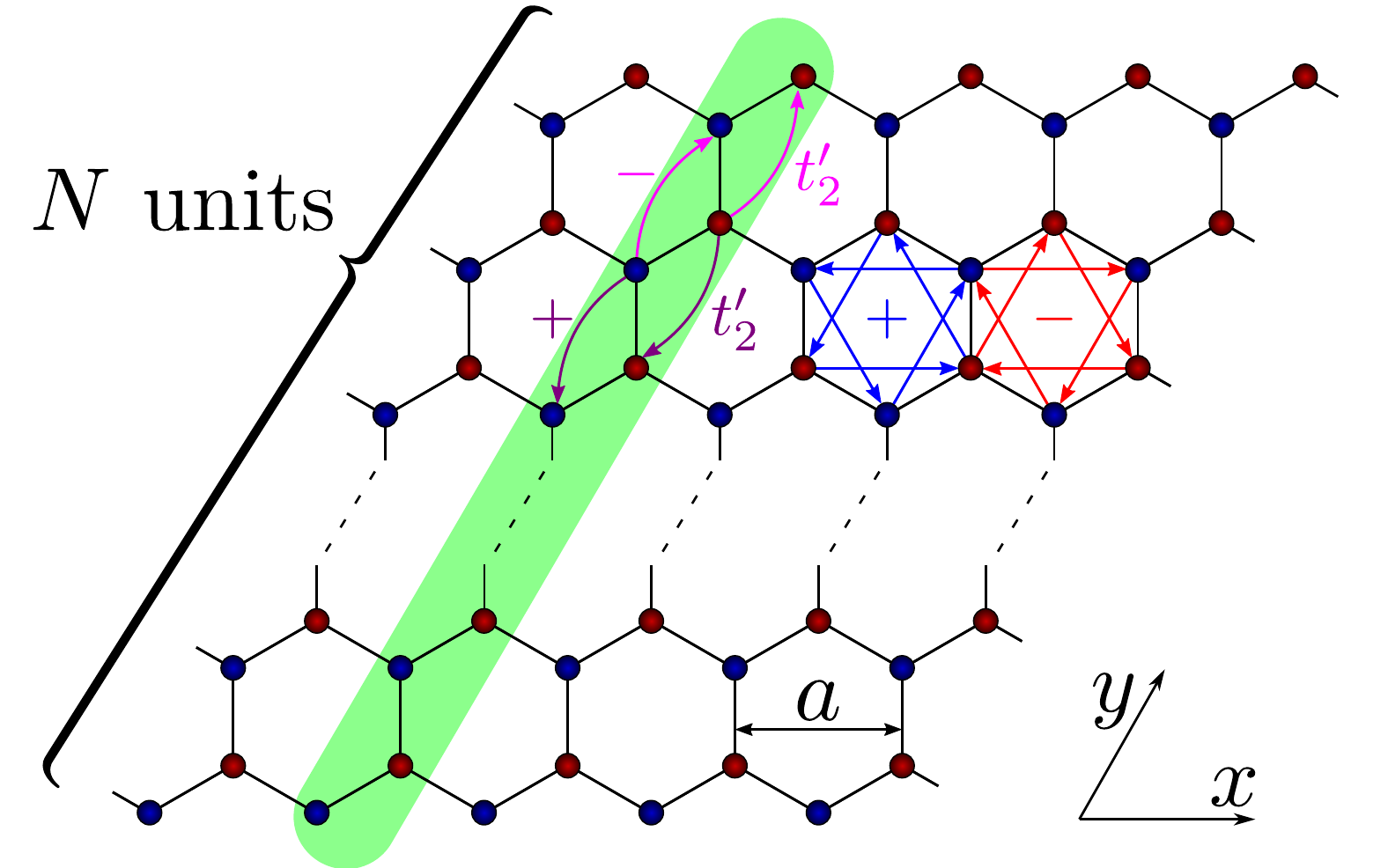}
	\caption{Infinite honeycomb strip in $x$-direction.
	NN hopping is depicted in black. A unit cell consists of $2 N$ sites in $y$-direction,
	shown in green. The sign of the phase in NNN hopping is
	given by arrows, e.g., red arrows stand for $- \phi$ or 
	magenta arrows for $+ \varphi$. The lattice constant $a$ is set to unity.}
	\label{fig:strip}
\end{figure}

The fermionic tight-binding model proposed by Haldane \cite{halda88b} as a first example
of non-trivial topological behavior without magnetic field is
a well-established model of a Chern insulator due to its simplicity. Hence, we choose it as 
our starting point. By adding a spatially anisotropic hopping it is possible to close
 the indirect gap while leaving the topological properties of the bands completely untouched.
 The Hamiltonian reads
\begin{subequations}
\label{eq:hamilton_haldane}
\begin{align}
\mathcal{H} =& \mathcal{H}_{\mathrm{Haldane}} + \mathcal{H}_{\mathrm{diag}} 
\\
\mathcal{H}_{\mathrm{Haldane}} =& t \sum_{\left\langle i, j \right\rangle} 
c_i^\dagger c_j^{\phantom{\dagger}} 
+ t_2 \sum_{\left\langle \left\langle i, j \right\rangle \right\rangle} 
e^{\pm \mathrm{i} \phi} c_i^\dagger c_j^{\phantom{\dagger}} 
\\
\mathcal{H}_{\mathrm{diag}} =& 
t_2' \sum_{\overline{\left\langle \left\langle i, j \right\rangle \right\rangle}} 
e^{\pm \mathrm{i} \varphi} c_i^\dagger c_j^{\phantom{\dagger}} ,
\end{align}
\end{subequations}
where $c_i^\dagger$ and $c_i^{\phantom{\dagger}}$ correspond to the creation and annihilation
 operators at \smash{site $i$,} respectively. The hoppings on the honeycomb lattice are shown in 
Fig.\ \ref{fig:strip}. A pair of nearest neighbor (NN) and next-nearest neighbor (NNN) 
sites is denoted by $\left\langle i, j \right\rangle$ and by 
$\left\langle \left\langle i, j \right\rangle \right\rangle$, respectively. 
The hopping elements $t, t_2$ and $t_2'$ are real and $t$ serves a energy unit.
The sign of the complex phase $\phi$ for the $t_2$-hopping is positive for 
anti-clockwise hopping and negative for clockwise hopping, see blue and red 
arrows in the plaquettes in \smash{Fig.\ \ref{fig:strip}.}

The notation $\overline{\left\langle \left\langle i, j \right\rangle \right\rangle}$ 
in the additional Hamiltonian $\mathcal{H}_{\mathrm{diag}}$ 
restricts the hopping to next-nearest neighbors in the $y$-direction. Therefore, it breaks the point group symmetry C3 of the bulk system. The sign of its phase $\varphi$ is positive
in $y$-direction and negative in $-y$-direction. This additional term may seem
artifical, but it is very suitable for the intended proof-of-principle.
Its realization in ultracold atom systems appears feasible \cite{jotzu14}.

In  reciprocal space the bulk Hamiltonian reduces to a $2\times2$
matrix due to the two sites in a unit cell; it can be expressed
in terms of Pauli matrices. One finds that $\mathcal{H}_{\mathrm{diag}}$ 
is given by $2 t_2' \cos(k_y + \varphi) \sigma_0$ where $\sigma_0$ is the identity matrix.
Hence the $t_2'$-hopping only induces an energy shift without having any
effect on the eigen states at given momentum. 
The topological properties derived from the eigen states such as the 
Berry curvature and the concomitant  Chern number \cite{fukui05} are preserved. 
The bulk dispersion, however, is altered due to $t_2'$.

\begin{figure}
	\centering
		\includegraphics[width=1.0\columnwidth]{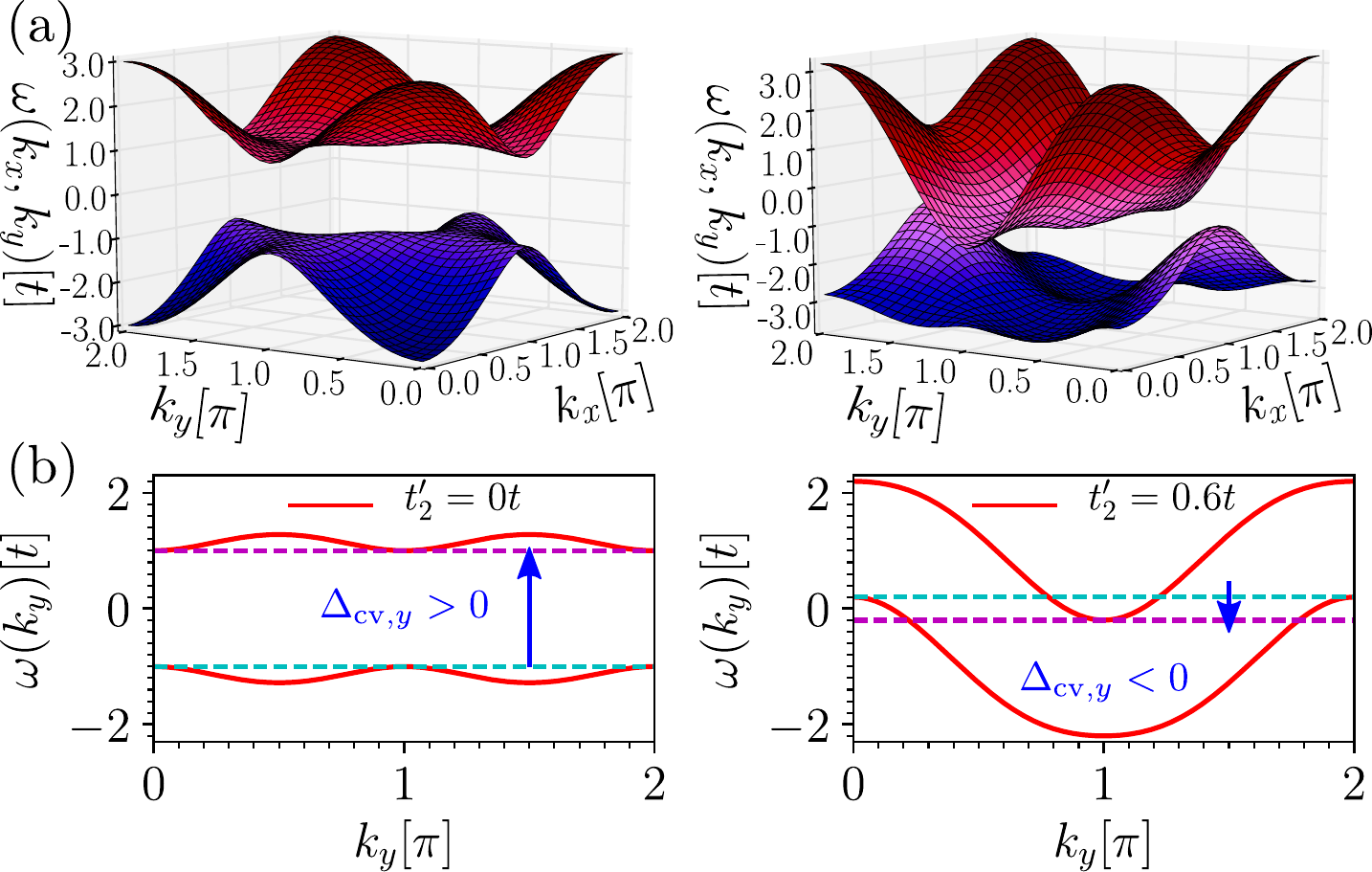} 
	\caption{(a) 
	Dispersion for $t_2 = 0.2 t$, $\phi = \pi/2$, $\varphi = 0$, and $t_2' = 0$ in
	the left panel and for $t_2' = 0.6 t$ in the right panel. (b) Dispersions as in (a) at 
	$k_x = \pi$. The magenta and cyan dotted lines indicate $\varepsilon_{\mathrm{bu,c}}$ and
	$\varepsilon_{\mathrm{bu,v}}$, respectively.}
	\label{fig:indirect}
\end{figure}

On the left hand side of Fig.\ \ref{fig:indirect} we illustrate the dispersion
for  $t_2 = 0.2 t$, $\phi = \pi/2$ without $t_2'$. If $t_2'$
is switched on (at $\varphi=0$) the dispersion changes significantly as shown
on the right hand side of Fig.\  \ref{fig:indirect}. The direct
energy gap at each given $k$-value does not change so that the two
bands stay well-separated. But the indirect gap 
is given by the energy difference between the green  and the blue dashed line
and hence vanishes and becomes even negative as displayed clearly in 
Fig.\ \ref{fig:indirect}(b) at fixed $k_x = \pi$.

To take the orientation of the boundary into account we define
the indirect gap $\Delta_{\mathrm{cv}, y}(k_x)$ as the smallest energy difference 
between the conduction and valence band at a fixed $k_x$, but for varied momentum $k_y$. 
The relevant band edge for the conduction band 
$\varepsilon_\mathrm{bu,c}(k_x) \coloneqq \mathrm{min}_{k_y} 
\omega_{\mathrm{bu, c}}(k_x, k_y)$ 
is displayed as green dotted line. For the valence band
 $\varepsilon_\mathrm{bu,v} (k_x) \coloneqq \mathrm{max}_{k_y}
\omega_{\mathrm{bu, v}}(k_x, k_y)$  
it is marked by the blue dotted line. Thus one has 
\begin{equation}
\Delta_{\mathrm{cv}, y}(k_x) = \varepsilon_{\mathrm{bu,c}}(k_x) 
- \varepsilon_{\mathrm{bu,v}}(k_x).
\end{equation}
This gap can take formally negative values. Tuning $t_2'$ from $0 t$ to $0.5 t$ at
 $\varphi = 0$ closes the indirect gap at $k_x = \pi$.

Next, we pass from the bulk to a finite, confined system considering a strip with zigzag edges as shown in Fig.\ \ref{fig:strip}. We investigate the existence of localized edge states.
The boundaries are chosen to run in $x$-direction and thus $k_x$ continues to be preserved,
but $k_y$ does not. Upon turning on the diagonal $t_2'$-hopping, the topological
 properties in the bulk remained completely unaffected, but we find a significant impact
on the system with boundaries: the exponentially localized edge states at $t_2'=0$
become less and less localized till they delocalize completely. We want to explore this
phenomenon here.

In order to measure the localization of states the inverse participation ratio 
\cite{krame93, calix15} (IPR) is most suitable. We want to quantify the localization 
to the edges of the strip, so we define the IPR of a normalized eigen state by
\begin{align}
I_n(k_x) =& \sum_i p_{n, i}^2(k_x) \nonumber \\
 =& \sum_i |\left\langle n, i | n, i\right\rangle|^2 (k_x) \quad \in [0, 1] \, ,
\end{align}
where $p_{n, i}$ is the probability of finding a particle at \smash{site $i$} 
in the unit cell in Fig.\ \ref{fig:strip} if the
system is in the $n$-th eigen state at momentum $k_x$. The IPR of localized states is finite, even for $N\to\infty$ while
it converges towards zero for delocalized, extended states in this limit.
Hence, in numerics an IPR of $O(1/N)$ indicates a delocalized state while
larger values indicate localization.

\begin{figure}
	\centering
		\includegraphics[width=1.0\columnwidth]{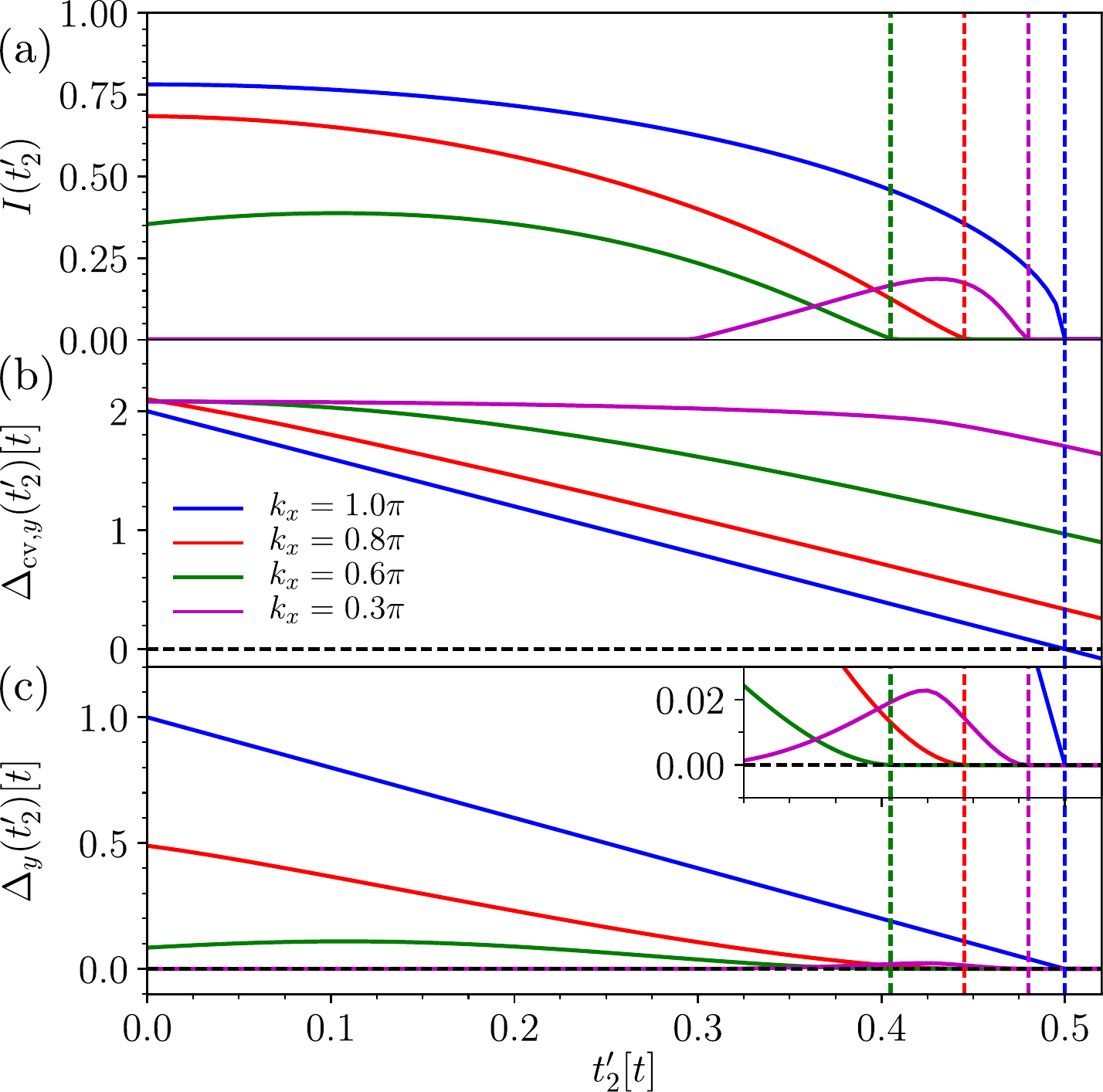} 
	\caption{(a-c) The IPR, $\Delta_{\mathrm{cv}, y}$, and $\Delta_y$ of the right-moving edge state vs.\ diagonal hopping $t_2'$ are shown for various momenta $k_x$ as computed 
	for $N=500$. $\Delta_y$ of both edge states at $k_x = 1.5 \pi$ lie on top of each other.}
	\label{fig:haldane_ipr}
\end{figure}

First, we focus on the case $k_x = \pi$ being the crossing point of the dispersion of
 the right and left moving in-gap state. Its energy lies precisely in the middle between
 the conduction and valence band rendering the spectrum at this value of $k_x$ similar to
the spectrum of the 1D case studied previously \cite{malki18b}.
Fig.\ \ref{fig:haldane_ipr}(a) depicts the IPR as a function of $t_2'$. 
For comparison, the indirect gap $\Delta_{\mathrm{cv}, y}$ is shown in Fig.\
 \ref{fig:haldane_ipr}(b). 
As in 1D, the IPR at $k_x = \pi$ decreases monotonically to its minimum value $O(1/N)$
upon increasing $t_2'$. The IPR reaches this value at the same value $t_2'$ where 
 the indirect gap $\Delta_{\mathrm{cv}, y}$ vanishes. This delocalized in-gap state remains 
extended for $\Delta_{\mathrm{cv}, y} < 0$. 

If $k_x$ takes other values the situation is more complex because the energy of the 
in-gap states is closer to one of the two bands, conduction or valence, respectively.
We observe that the delocalization $I\approx 0$ occurs for smaller values of $t_2'$
than the zero of the indirect gap $\Delta_{\mathrm{cv}, y}$, see 
Fig.\ \ref{fig:haldane_ipr}(a) and (b).
So we conclude that existence of an indirect gap and delocalization are linked, 
but not in a straightforward manner, see discussion below.

In order to achieve a better understanding we define a specific indirect gap $\Delta_y$
referring to the energy of the in-gap state. This piece of information is available
once the strip geometry is analyzed quantitatively. Let the in-gap energies be denoted
by  $\omega_{\mathrm{in},\alpha}$ where $\alpha$ denotes the different in-gap branches. 
Then $\Delta_y$ is the smallest energy difference of $\omega_{\mathrm{in},\alpha}$
to one of the bands at fixed $k_x$
\begin{align}
\Delta_y(k_x,\alpha) := \mathrm{min} \left\lbrace \omega_{\mathrm{in},\alpha} - \varepsilon_\mathrm{bu,v}, 
\varepsilon_\mathrm{bu,c} - \omega_{\mathrm{in},\alpha} \right\rbrace .
\end{align}
If the in-gap states enter the continua of either conduction or valence band we set 
$\Delta_y(k_x,\alpha)=0$. Thus, $\Delta_y(k_x,\alpha)=0$ measures the energy distance of in-gap states 
to the extended bulk modes. It is to be expected that it is closely related to delocalization.

The indirect gap $\Delta_y$ as function of $t_2'$ is shown in \smash{Fig.\ \ref{fig:haldane_ipr} (c).} 
For $k_x = \pi$, $\Delta_y$ behaves like $\Delta_{\mathrm{cv}, y}$ since in this particular symmetric case 
both quantities are proportional to each other. For other momenta, however, differences appear.
 In contrast to $\Delta_{\mathrm{cv}, y}$, $\Delta_y$ at $k_x \ne \pi$ vanishes exactly at
the value of $t_2'$ where the IPR essentially vanishes. 
This shows that localization can be attributed to a finite $\Delta_y$. 
Note also the possible non-monotonic behavior of IPR and $\Delta_y$ as function
of $t_2'$, e.g., at $k_x = 0.3 \pi$.

\begin{figure}
	\centering
		\includegraphics[width=1.0\columnwidth]{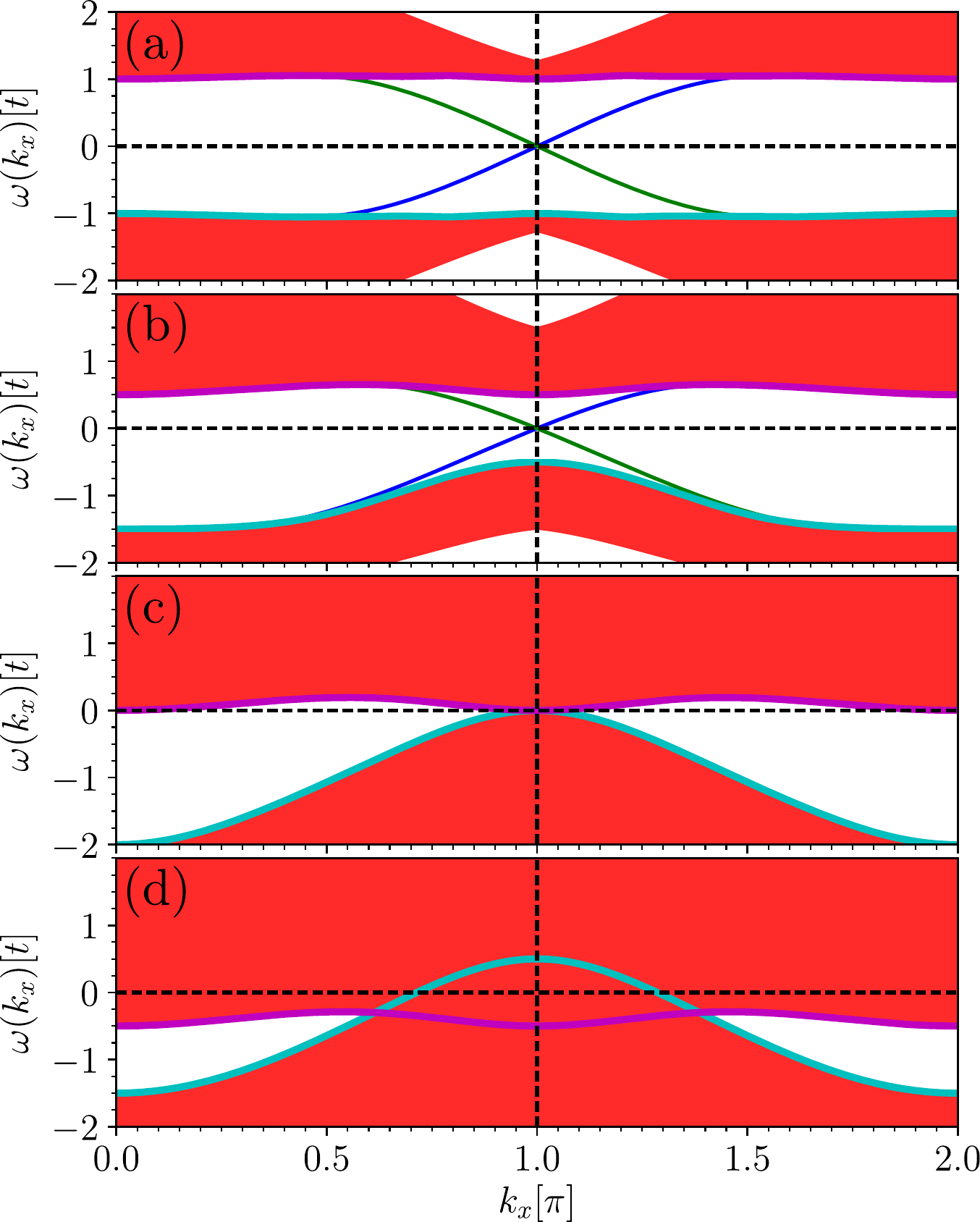} 
	\caption{(a-d) Continua of the two bulk bands and dispersions of
the two in-gap states (right-mover in blue, left-mover in green) for $t_2 = 0.2 t$, $\phi = \pi/2$, $\varphi = 0$, and $t_2' = \left\lbrace 0 t, 0.25 t, 0.5 t, 0.75 t \right\rbrace$. Due to the absence of an indirect gap the continua overlap in panels (c) and (d) and no in-gap states can be identified. The magenta and cyan lines indicates the band edges $\varepsilon_{\mathrm{bu,c}}$ and $\varepsilon_{\mathrm{bu,v}}$, respectively.}
	\label{fig:haldane_evolve}
\end{figure}

For the sake of comprehensibility we visualize
the evolution of the band structure as a function of the hopping
amplitude $t_2'$.
In \smash{Fig.\ \ref{fig:haldane_evolve}} we depict four representative cases 
$t_2'= \left\lbrace 0t, 0.25t, 0.5t, 0.75t \right\rbrace$. On increasing $t_2'$
the conduction and valence bulk bands are approaching each
other and the edge states are becoming covered by them more and more,
see Fig.\ \ref{fig:haldane_evolve}(a) and (b). At the marginal value $t_2'=0.5t$
shown in \smash{Fig.\ \ref{fig:haldane_evolve}(c)}, all in-gap states are covered by bulk states and therefore are
delocalized. This coincides with the closing of the indirect gap 
$\Delta_{\text{cv}, y}=0$ at $k_x = \pi$. Increasing $t_2'$ further,
see  \smash{Fig.\ \ref{fig:haldane_evolve}(d),} the range of $k_x$-values increases where $\Delta_{\text{cv},y}$
is zero or negative.

There is a large number of further aspects worth investigating: (i) In the Supplement
\cite{supplement} we study the case $\varphi=\pi/2$ which confirms our 
conclusion that the vanishing of the indirect gap $\Delta_y$ goes along
with delocalized in-gap states. But better localized states may have a smaller
$\Delta_y$ which shows that both quantities are not linked by a simple
monotonic relation. (ii) We find 
that if the additional hopping runs along $x$ and not along $y$  the additional term
reads $2 t_2' \cos(k_x + \varphi) \sigma_0$ and does neither change the bulk topology
nor the localization in the strip in Fig.\ \ref{fig:strip}.
(iii) Samples which are finite in both directions
are also studied \cite{supplement}. We find that their chiral edge states
become extended precisely if along one of the edges the in-gap states delocalize.
Finally, we point out that  different boundaries imply
 different edge states dispersions. For instance, a bearded boundary \cite{uhrig16} 
in the Haldane model has its crossing point a $k_x = 0$ implying a different $\Delta_y(k_x)$
so that the localization persists up to larger values of $t_2'$.

\begin{figure}
	\centering
		\includegraphics[width=1.0\columnwidth]{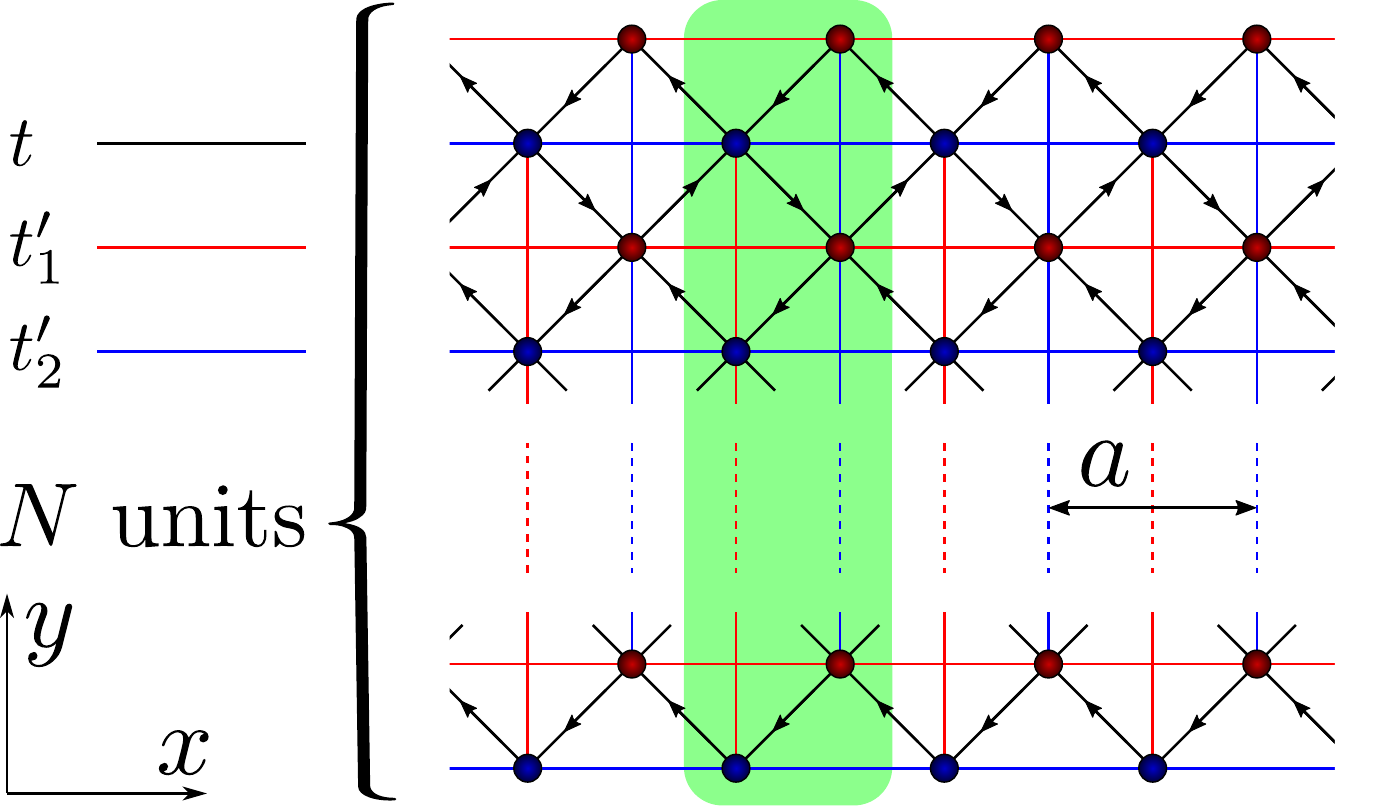} 
	\caption{Checkerboard strip with NN hopping (black bonds). Hopping in direction of the arrows have a positive
	sign. The red (blue) line stands for $t_1'$ ($t_2'$) hopping. The lattice constant $a$ is set to unity.}
	\label{fig:strip_square}
\end{figure}

\begin{figure}
	\centering
		\includegraphics[width=1.0\columnwidth]{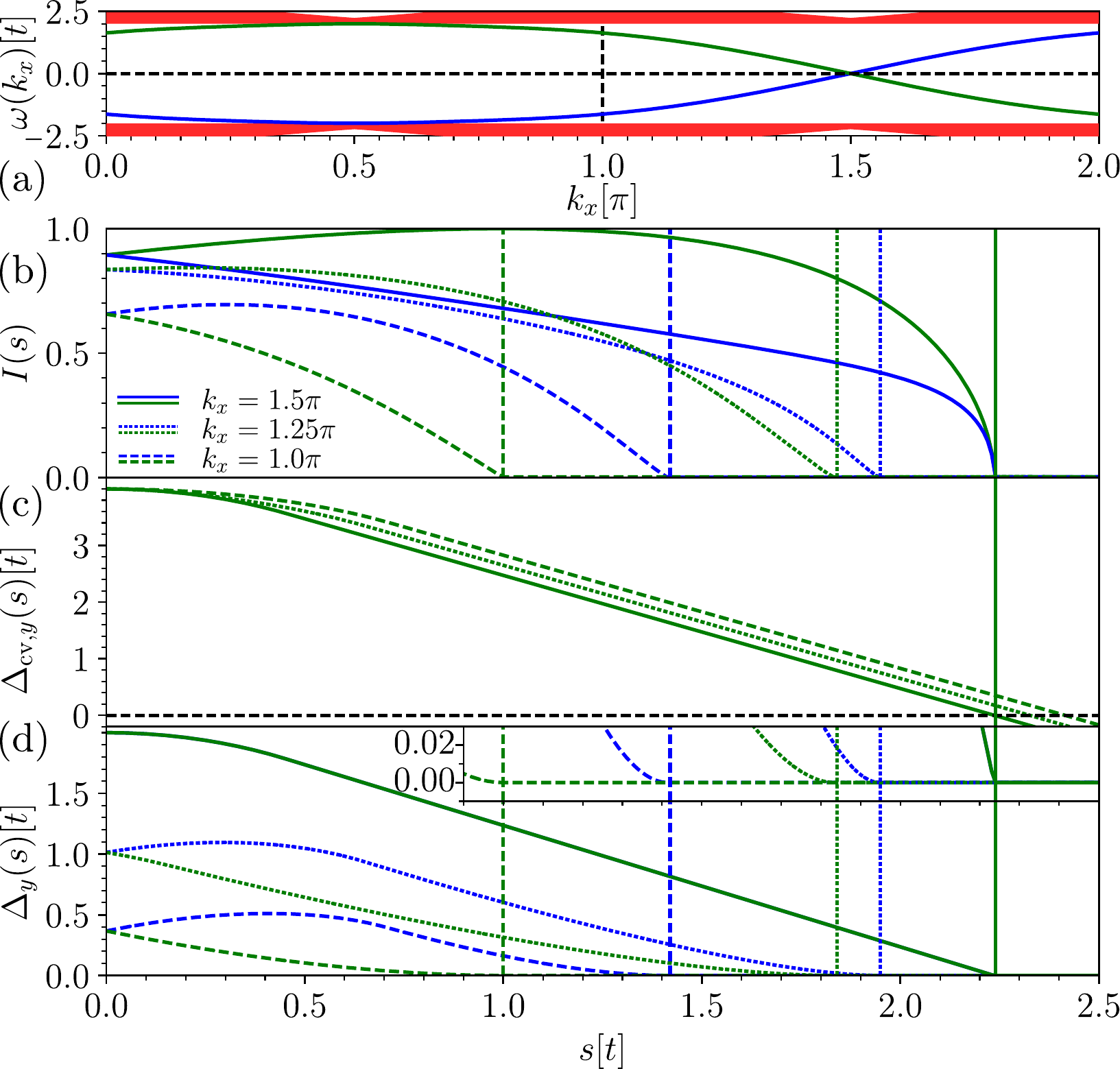}
	\caption{(a) Dispersions for $s = 0 t$, 
	$d = -1 t$, and $\phi = \pi/4$. The continua of the bulk bands are shown as red 
	areas. (b-d) The IPR, $\Delta_{\mathrm{cv}, y}$, and $\Delta_y$ of the right-moving 
	and left-moving edge state vs.\ the parameter $s$ are shown for various wave vectors
	$k_x$ as computed for $N=500$.}
	\label{fig:square}
\end{figure}

The standard lattice studied above has provided a proof-of-principle result allowing
us to establish the importance of indirect gaps for the localization of in-gap states so
that they represent true edge states. In order to corroborate that this scenario is
generic and experimentally relevant we next address the topological checkerboard lattice,
see Fig.\ \ref{fig:strip_square}, which 
has been realized by optical lattices  \cite{olsch12, aidel13, miyak13}.
This lattice is
is described by a two-band model \cite{sun11} with NN ($t$) and NNN ($t_1'$, $t_2'$) 
hopping
\begin{align}
\mathcal{H} = - t \sum_{\left\langle i, j \right\rangle} e^{\pm \mathrm{i} \phi} c_i^\dagger c_j^{\phantom{\dagger}} - 
\sum_{\left\langle \left\langle i, j \right\rangle \right\rangle} t_{ij}' c_i^\dagger c_j^{\phantom{\dagger}}. 
\end{align}
For the bulk, Fourier transformation yields a representation
in terms of Pauli matrices
\begin{align}
\mathcal{H} =& - s (\cos(k_x) + \cos(k_y)) \sigma_0 \nonumber  \\
 &- d (\cos(k_x) - \cos(k_y)) \sigma_z \nonumber \\
 &- 4 t \cos(\phi) \cos(k_x/2)\cos(k_y/2) \sigma_x \nonumber \\
 &- 4 t \sin(\phi) \sin(k_x/2)\sin(k_y/2) \sigma_y 
\end{align}
where we use $s:=t_1'+t_2'$ and $d:=t_1'-t_2'$ for brevity and $t$ as energy unit. 
A topological phase occurs for $\phi \neq n \pi$ and $d \neq 0$ \cite{sun11}.
Investigating the strip sketched in \smash{Fig.\ \ref{fig:strip_square}} one clearly sees the left and right
moving in-gap states shown in panel (a) of Fig.\ \ref{fig:square}. 
Tuning $s$ while keeping $d$ constant \cite{supplement}, the bulk topology
is not changed, but the dispersion changes, just as for the Hamiltonian \eqref{eq:hamilton_haldane}.
Indeed, we find the same scenario as in Fig.\ \ref{fig:haldane_ipr}, 
see panels (b) to (d) in \smash{Fig.\ \ref{fig:square}.}
This strongly corroborates our findings and paves the way to their experimental 
verification. 

Summarizing, non-trivial topological properties of the bulk imply the existence
of in-gap states. Often, they are supposed to be localized at the 
boundaries of the sample. But in generic one-particle models we showed that 
these edge states can delocalize if they are not protected by finite indirect gaps.
Mostly clearly, this can be demonstrated by adding terms to the Hamiltonians proportional 
to the identity matrix. They change the dispersions, but leave the eigen states unchanged and
hence the topological properties. We stress that this holds true independent
of the number of bands. This message also implies that the omission of terms
proportional to the identity matrix is acceptable for the bulk, but not 
for confined geometries.

For in-gap states of which the energy is protected by additional symmetries it  is 
sufficient to consider the bulk indirect gap $\Delta_{\text{cv},y}$. Generally, this gap
is not sufficient to decide on localization and one has to consider the indirect gap $\Delta_y$ which measures
the energetic distance of the in-gap states to the closest bulk band. 
Generically, if $\Delta_y$ is finite the states are localized and thus true edge states.
If $\Delta_y$ vanishes delocalization is to be expected. 

While the described scenario is the generic one it can vary in special cases.
Baum and co-workers \cite{baum15} pointed out that further symmetries such
as momentum and energy conservation can prevent delocalization in topological states
of matter in spite of coupling edge states to a gapless bulk. Similarly, Verresen and co-workers
\cite{verre18a} discovered edge states at the ends of critical chains.
Independent of topological properties, it has been noted that 
localization can persist notwithstanding hybridization with continua in 
especially designed systems \cite{molin12}. The localization may be weak in
the sense that it is not exponential, but algebraic \cite{corri13}.

Yet, the results presented in this Letter for standard one-particle topological models 
illustrate that delocalization of edge states is the generic phenomenon if indirect gaps
vanish and hybridization with bulk continua occurs. To the best of
our knowledge, this fact has not yet been appreciated in literature even though
it has important consequences for realizations of topological phases 
and their experimental detection. The take-home message is that the lack of localized
edge modes does not preclude the existence of non-trivial topology 
characterized by discrete topological invariants. Then, however, direct techniques to
detect topological invariants are \smash{required \cite{atala13,zeune15,mitta16}.}

To pave the way towards experimental verifications by ultracold atoms in
optical lattices we considered the topological checkerboard model explicitly.
Further preliminary results show that the advocated scenario also
occurs in the Kane-Mele model including Rashba couplings as a prototypical model with 
$\mathbb{Z}_2$-topological invariant.

\paragraph{Acknowledgments}

This work was supported by the Deutsche Forschungsgemeinschaft and the Russian Foundation of Basic Research 
in TRR 160. MM gratefully acknowledges 
financial support by the Studienstiftung des deutschen Volkes. GSU thanks Oleg P.\
Sushkov for useful discussions and the School of Physics of the 
University of New South Wales for its hospitality and  
the Heinrich-Hertz Foundation for financial support of this visit.

%\bibliographystyle{eplbib}
%\bibliographystyle{apsrev4-1}
%\bibliography{liter10}


%%%%%%%%%% Merge with supplemental materials %%%%%%%%%%
%\pagebreak
\clearpage
\pagebreak

%\widetext
%\begin{center}
%\textbf{\large Supplementary Materials}
%\end{center}
%%%%%%%%%% Merge with supplemental materials %%%%%%%%%%
%%%%%%%%%% Prefix a "S" to all equations, figures, tables and reset the counter %%%%%%%%%%
\setcounter{equation}{0}
\setcounter{figure}{0}
\setcounter{table}{0}
\setcounter{page}{1}
\makeatletter
\renewcommand{\theequation}{S\arabic{equation}}
\renewcommand{\thefigure}{S\arabic{figure}}
\renewcommand{\bibnumfmt}[1]{[S#1]}
\renewcommand{\citenumfont}[1]{S#1}
%%%%%%%%%% Prefix a "S" to all equations, figures, tables and reset the counter %%%%%%%%%%

\vspace*{0.5cm}

\onecolumngrid
\begin{center}
\textbf{\large  Supplemental Material for "Delocalization of edge states in topological phases"}
\end{center}
\twocolumngrid

\section{Delocalization induced by an additional kinetic term} 

\begin{figure}[htb]
	\centering
		\includegraphics[width=1.0\columnwidth]{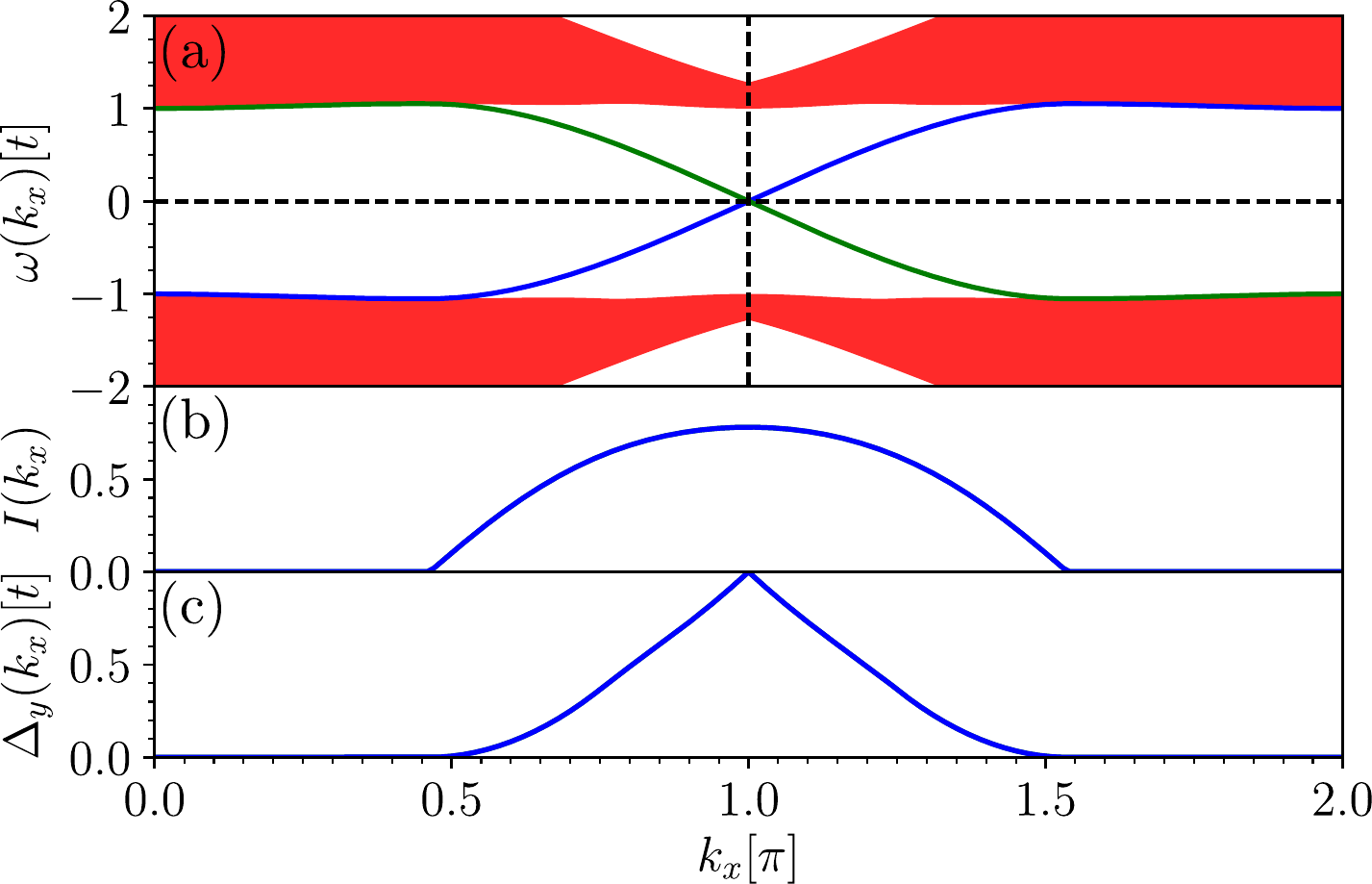}
	\caption{(a) Continua of the two bulk bands and dispersions of the in-gap states 
	(right-mover in blue, left-mover in green) for $t_2 = 0.2 t$, $\phi = \pi/2$, and $t_2' = 0$. 
	The continua of the bulk bands are marked by filled red areas. 
	(b) The IPR of both edge states as function of wave vector $k_x$. 
	(c) The indirect gap of the edge states $\Delta_y$ vs.\ wave vector $k_x$.}
	\label{fig:haldane_origin}
\end{figure}

Here we study the effect of the additional
diagonal hopping term on the (de)localization of edge states in more detail.
The dispersion of the original Haldane model as given in Eq.\ (1) in the main text
in a strip geometry with zigzag edge is shown in 
Fig.\ \ref{fig:haldane_origin} (a). The two dispersion branches marked in blue and green 
are connecting the valence and conduction band. They belong to the right and left-moving in-gap states 
with energies $\omega_{\mathrm{in},\alpha}$ where $\alpha$ labels the two branches. 
In the same range of parameters where the energy of the in-gap states is clearly distinct
from the bulk continua (shown in red) the inverse participation ratio (IPR) \cite{krame93,calix15}
is finite indicating well localized edge states, see Fig. \ref{fig:haldane_origin}(b). 
The energy separation of the in-gap states from the closest bulk energies is described 
by the specific indirect gap $\Delta_y$ defined in Eq.\ (4) in the main text.
It is displayed in Fig. \ref{fig:haldane_origin}(c). In all three panels, the blue curve
refers to the right-mover and the green curve to the left mover.
Clearly, a finite value of $\Delta_y$ and a finite value of the IPR go along with each other.
Hence the corresponding in-gap states are truly localized edge states, one at the top and one at the bottom
of the strip. Due to reflection symmetry both edge states show the same IPR dependence. 
As a result the blue and green curves in Fig.\ \ref{fig:haldane_origin}(b) and (c) lie on top
of each other. We see that the IPR increases for increasing $\Delta_y$ upon variation of $k_x$. 

\begin{figure}[htb]
	\centering
		\includegraphics[width=1.0\columnwidth]{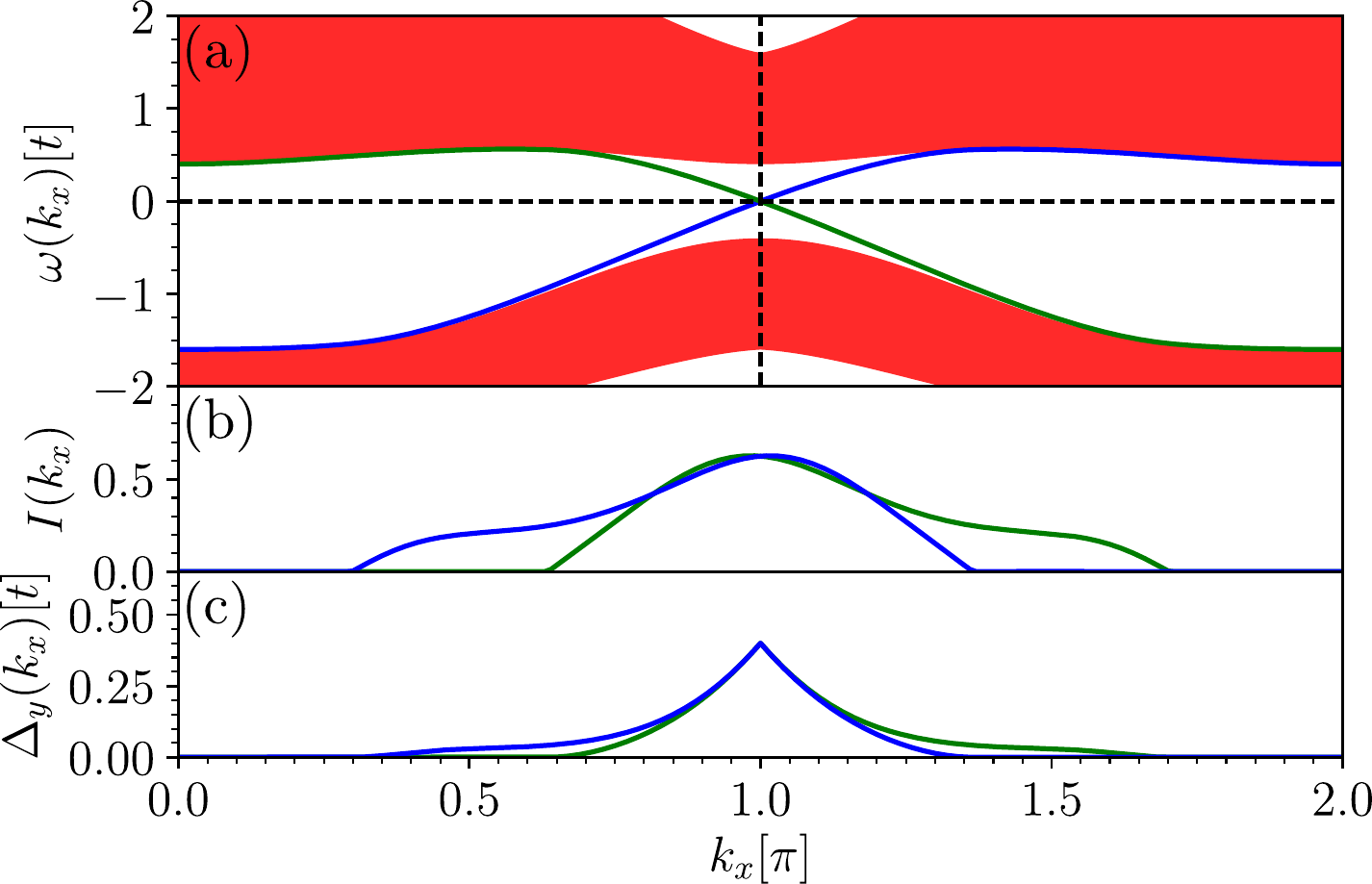}
	\caption{(a) Continua of the two bulk bands and dispersions of the two in-gap states 
	(right-mover in blue, left-mover in green) for $t_2 = 0.2 t$, $\phi = \pi/2$, $t_2' = 0.3 t$,
	and $\varphi = 0$, i.e., real NNN hopping. The continua of the bulk bands are marked by filled red areas. 
	(b) The IPR of both edge states as function of wave vector $k_x$. 
	(c) The indirect gap of the edge states $\Delta_y$ vs.\ wave vector $k_x$.}
	\label{fig:haldane_real}
\end{figure}

\begin{figure}
	\centering
		\includegraphics[width=1.0\columnwidth]{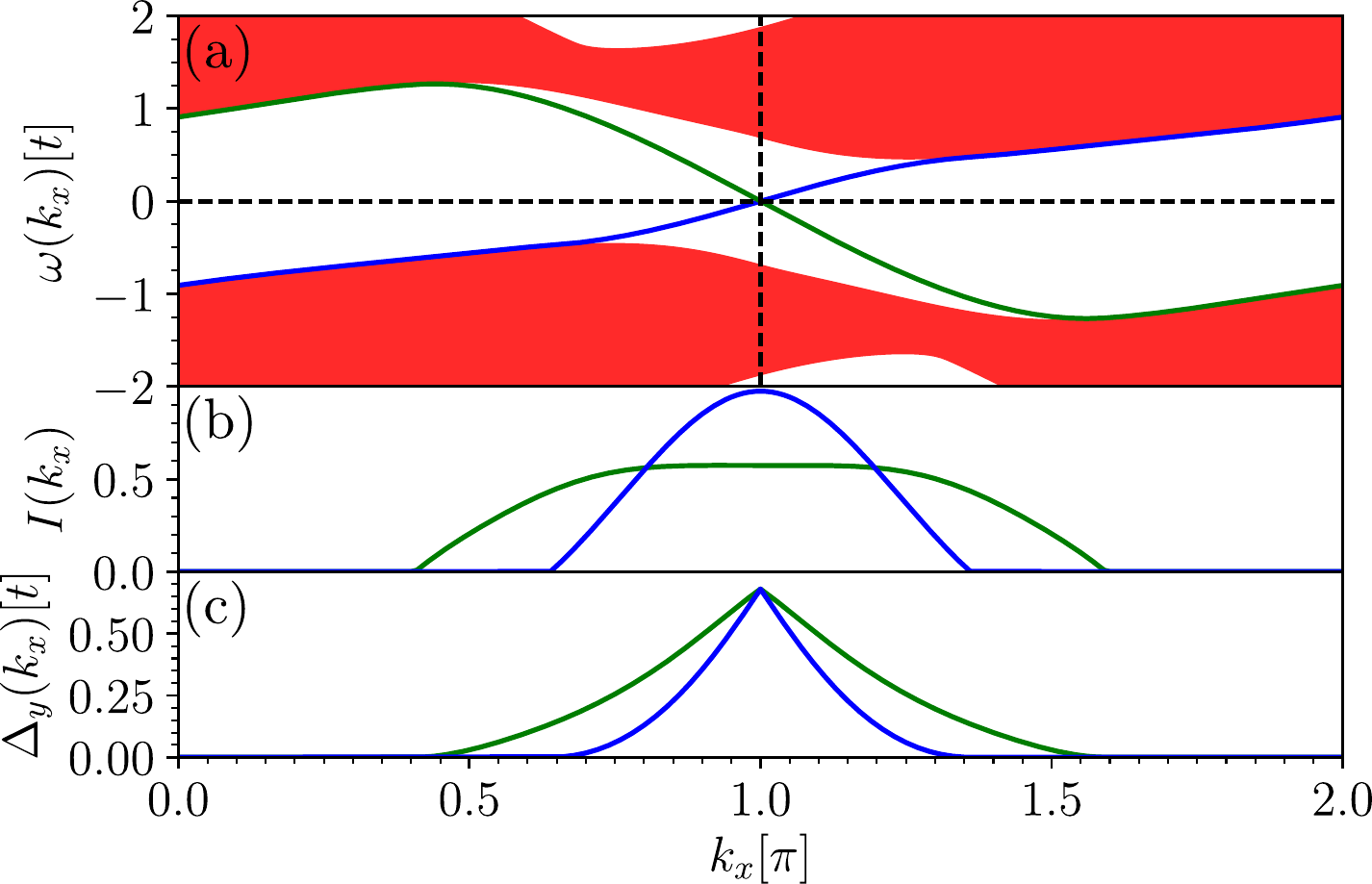}
	\caption{(a) Continua of the two bulk bands and dispersions of the two in-gap states 
	(right-mover in blue, left-mover in green) for $t_2 = 0.2 t$, $\phi = \pi/2$, $t_2' = 0.3 t$, and 
	$\varphi = \pi/2$, i.e., imaginary NNN hopping. The continua of the bulk bands are marked by filled red areas. 
	(b) The IPR of both edge states as function of wave vector $k_x$. 
	(c) The indirect gap of the edge states $\Delta_y$ vs.\ wave vector $k_x$.}
		\label{fig:haldane_imag}
\end{figure}

Next, we  study the effect of changing the indirect gap by turning on $t_2'$ for real hopping, i.e.,
for $\varphi = 0$, \smash{see Fig.\ \ref{fig:haldane_real},} and for imaginary hopping, i.e., $\varphi = \pi/2$,
see Fig.\ \ref{fig:haldane_imag}. Since $t_2'$ breaks the particle-hole symmetry the left and right moving edge state differ from each other for $t_2' \neq 0$. 

Fig.\ \ref{fig:haldane_real}(a) depicts exemplary results which show that the conduction band edge
is lowered such that the indirect gap $\Delta_y$ and the IPR vanish earlier for the right-movers
for $k_x > \pi$ and for the left movers for $k_x < \pi$.
In contrast, the valence band is lowered such that the energy range for distinct
 edge states is increased. Thus, $\Delta_y$ becomes finite in additional regions, namely 
for smaller $k_x$ for the right-movers and for larger $k_x$ for the left-movers.
This is particularly evident in comparison to  Fig.\ \ref{fig:haldane_origin}. 
As consequence, the curves for the IPR and for the indirect gaps
no longer have axial symmetry about $k_x = \pi$ or $k_x = 0$, see Fig.\ \ref{fig:haldane_real}(b) and (c).
But reflection about one of these axes interchanges right- and left movers. Consequently, the localization analysis of one edge state as shown in \smash{Fig.\ 3} of the main text is sufficient. 

For completeness, we illustrate the delocalization of edge states as a result of imaginary diagonal hopping
for $\varphi = \pi/2$. This hopping alters the edges of the bulk continua considerably
spoiling their axial symmetry. The dispersions and the bulk edges are 
inversion symmetric with respect to $(\pi,0)$ as can be seen in Fig.\ \ref{fig:haldane_imag}(a). Thus,
the IPR of an edge state is axial symmetric with respect to $k_x = \pi$ or $k_x = 0$. 
As for the case of real hopping,
only a finite indirect gap $\Delta_y$ yields a finite value of the IPR 
in the thermodynamic limit $N\to\infty$.
We point out that the imaginary hopping has a different impact on the localization than the real hopping.
 For instance, the IPRs for the edge states at $k_x = \pi$ are different 
while their indirect gaps are the same. Hence it is clear that there is no general 
relation between both quantities. Of course, this was to be expected since the IPR is dimensionless 
while the indirect gap has the unit of an energy. 
Clearly, a velocity $v$ and the lattice constant $a$
must enter at least in a quantitative relation between IPR and $\Delta_y$.

\begin{figure}[h]
	\centering
		\includegraphics[width=1.0\columnwidth]{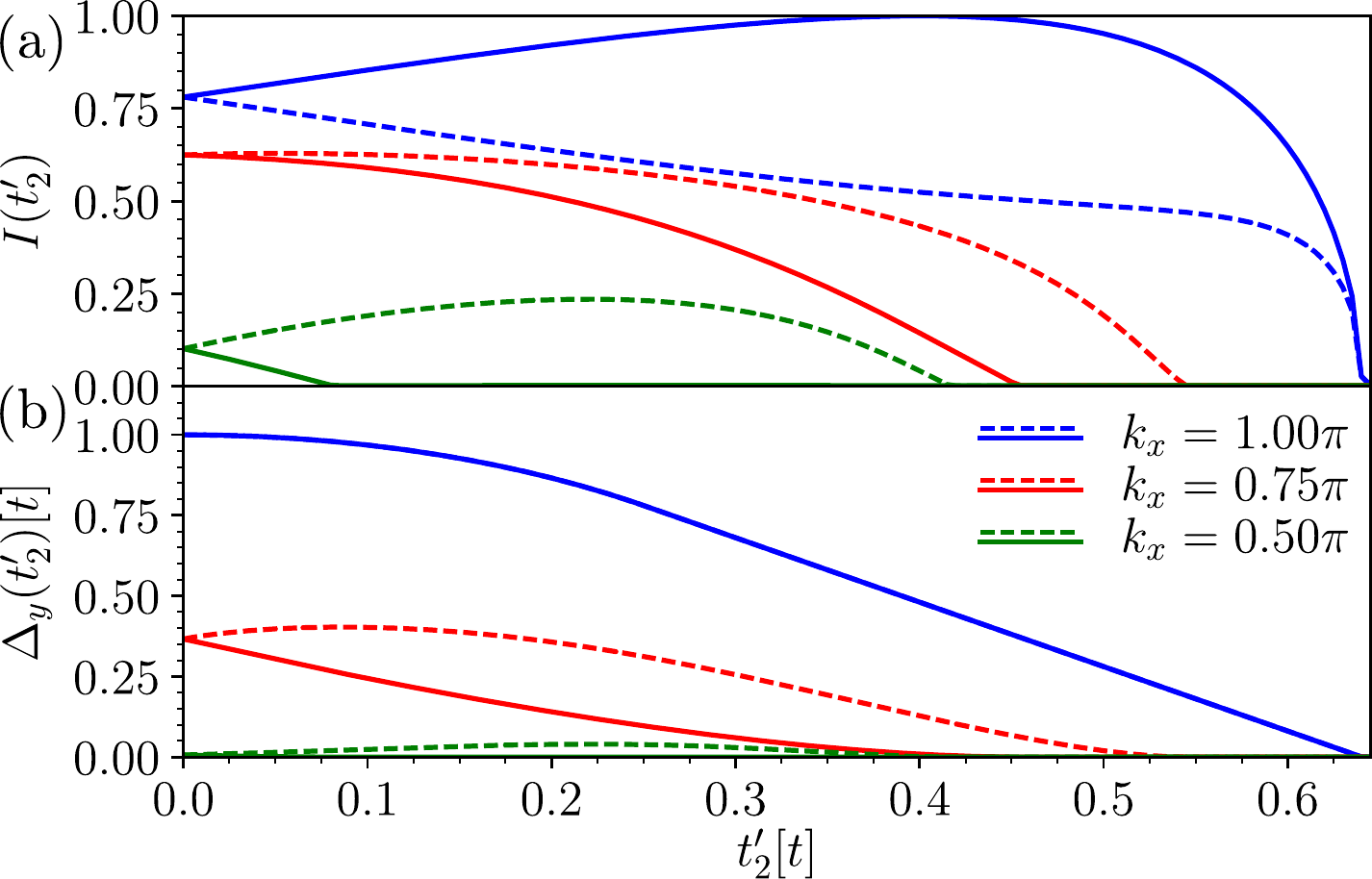}
	\caption{(a - b) The IPR and $\Delta_y$ of both edge state vs.\ hopping parameter $t_2'$ for 
	various wave vectors $k_x$. The values of the right-moving edge state are shown as solid line 
	while the dashed line belongs to the left-moving edge state. For $\Delta_y$ at $k_x=\pi$,
	they lie on top of each other.}
	\label{fig:haldane_imag2}
\end{figure}

Due to the broken reflection symmetry of the dispersion the two edge states display different dependencies. 
The IPR of the edge states as function of $t_2'$ is shown in Fig.\ \ref{fig:haldane_imag2}. Inspecting
 the IPR of the right-moving edge state at $k_x = \pi$ one discerns that the IPR first increases 
for increasing $t_2'$ despite the decrease of the indirect \smash{gap $\Delta_y$.} Thus, it is 
corroborated that  the localization does not only depend on the indirect gap $\Delta_y$. But just as in the 
case of real hopping the vanishing of the indirect gap induces  delocalization. 
Note that the eigen states at $k_x = \pi$ are doubly degenerated; nonetheless their IPRs are different. 
In addition, the IPRs of both edge states depending $t_2'$ are presented in Fig.\ \ref{fig:haldane_imag2}. 
Qualitatively, the relation between the IPRs and the indirect gaps $\Delta_y$ are similar to the case
of real hopping.

\section{Delocalization of chiral edge states}

Edge modes are mostly considered and computed for infinite strip geometries because they
allow one to consider models which preserve one translational symmetry. The edge modes can be identified 
easily by looking for gapless dispersion branches between two bulk bands. 
For finite samples which are confined in all directions the analysis becomes much more
intricate because the lack of any momentum conservation makes it difficult to 
identify the energies of edge modes in the energy spectrum. 

A possible solution is to deduce the indirect energy gap in the bulk allowing
for changes of all wave vectors if it is finite. Energies of the finite sample 
lying within the energy window given by the finite indirect gap
are associated to edge modes.
This method can be used for topological insulators with appropriate finite
indirect gaps, but it fails if the indirect gap closes or if
 the system even enters the phase of a topological metal \cite{ying18}. 

\begin{figure}
	\centering
		\includegraphics[width=1.0\columnwidth]{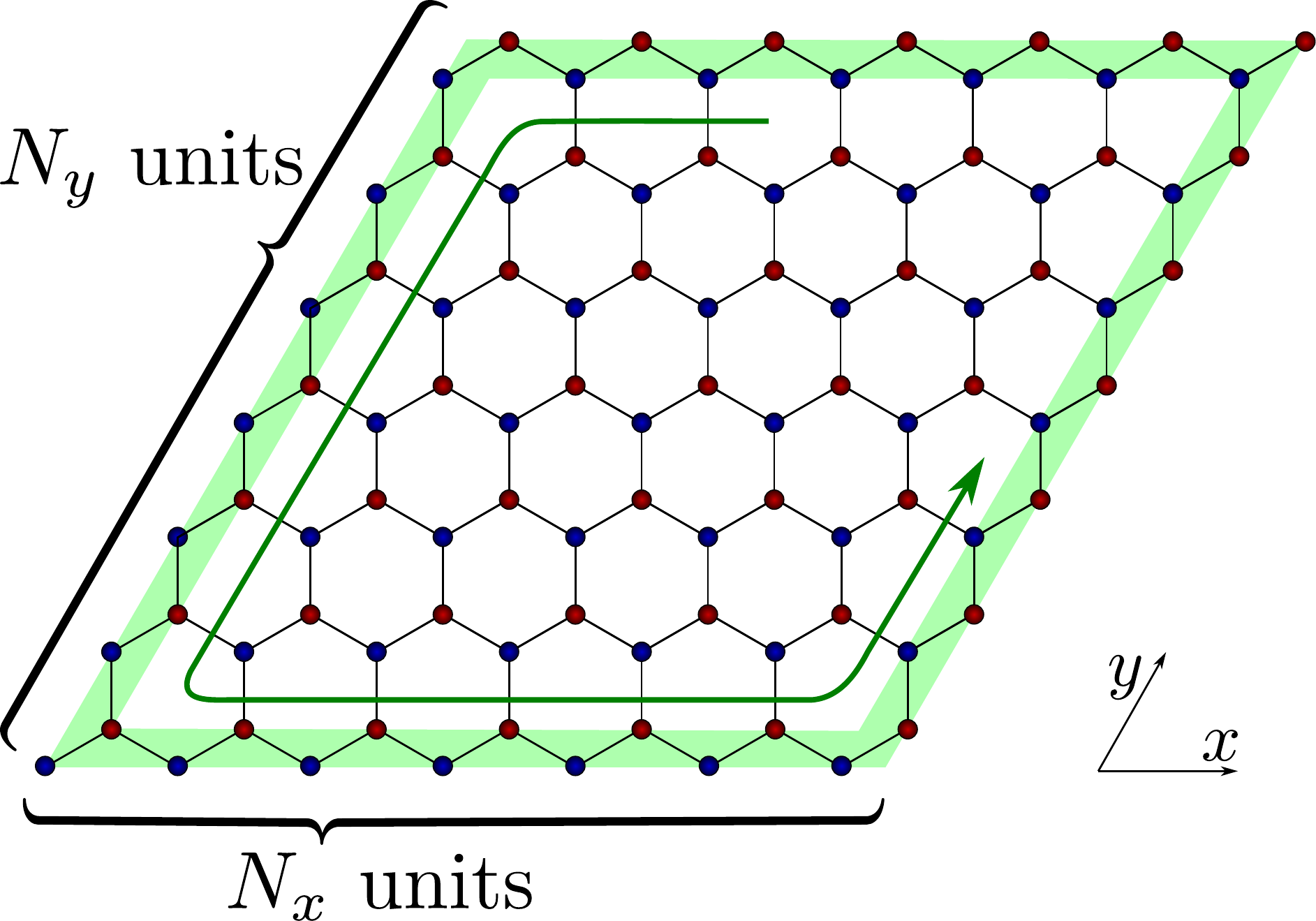}
	\caption{Sketch of a finite sample geometry. The localization area of a chiral edge state 
	is highlighted in green. A possible orientation of the chirality 
	is indicated by the green arrow.}
	\label{fig:sample}
\end{figure}

The edge mode in a finite sample is localized along the entire boundary and the particle in such
a state is propagating only in one direction as shown in Fig.\ \ref{fig:sample}. 
Therefore, such edge modes are called chiral edge mode. 
For large samples, the number of sites close to the boundary relative to the total number $N_\text{tot}$ of
sites is small and tends to zero for $N_\text{tot}\to\infty$.
This fact opens the possibility to identify edge modes by their IPR:
the states with the largest IPRs are the best localized ones which are to be found
along the boundary. (Note, however, that  this approach does not work for disordered
samples where fully localized states may exist in the bulk.)

\begin{figure}
	\centering
		\includegraphics[width=0.93\columnwidth]{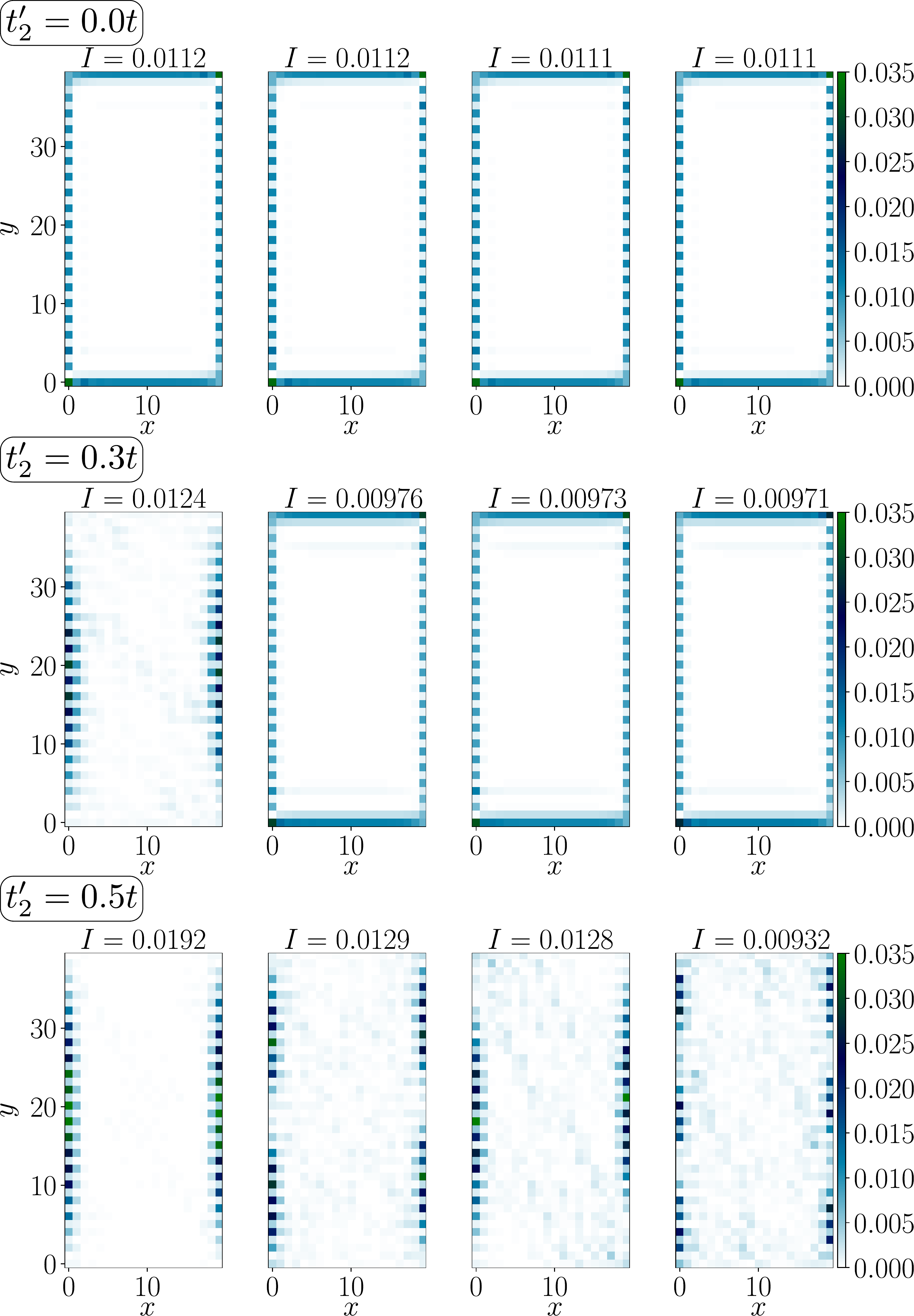}
	\caption{Probabilities in a 2D sample with $2 N_x \times N_y = 2\cdot 20\times 20 = 800$ sites.
	The four eigen states with the largest IPRs are depicted in each row at
	$t_2 = 0.2 t, \phi = \pi/2, \varphi = 0$ for $t_2' = \left\lbrace 0t, 0.3t, 0.5 t \right\rbrace$, respectively. }
	\label{fig:psi}
\end{figure}

Here, the IPR of an eigen state is defined as
\begin{subequations}
\begin{align}
I_n =& \sum_i p_{n, i}^2 \\
 =& \sum_i |\left\langle n, i | n, i\right\rangle|^2 \quad \in [0, 1] 
\end{align}
\end{subequations}
where the sum runs over \emph{all} sites of the sample.
\smash{Fig.\ \ref{fig:psi}} depicts the probabilities of the four eigen states with the highest IPRs for 
$t_2' = \left\lbrace 0t, 0.3t, 0.5 t \right\rbrace$ in the finite 2D sample. 
The case $t_2' = 0$ corresponds to the original Haldane model with its known topological characteristics. 
As expected, all the four eigen states display localization along the complete boundary indicating that
they are indeed chiral states. Increasing $t_2'$ implies that less and less eigen states 
show finite probabilities along the complete boundary. 
But as long as there is a finite indirect gap between the
 conduction and the valence band  chiral edge states exist.  

At $t_2' = 0.5 t$, the indirect gap has vanished, see also main text. Indeed, no chiral edge states 
can be found anymore. The displayed eigen states in Fig.\ \ref{fig:psi} are localized at edges running
along $y$-direction because this localization is not altered by the diagonal hopping $t_2'$ as we 
observed already in the main text (with the roles of $x$ and $y$ interchanged).
But we stress that the localization at the edges running in $x$-direction is completely eradicated 
due to the diagonal hopping $t_2'$ as expected from the calculations for strip geometry in the
main text.

\section{Delocalization in the topological checkerboard model}

\begin{figure}
	\centering
		\includegraphics[width=1.0\columnwidth]{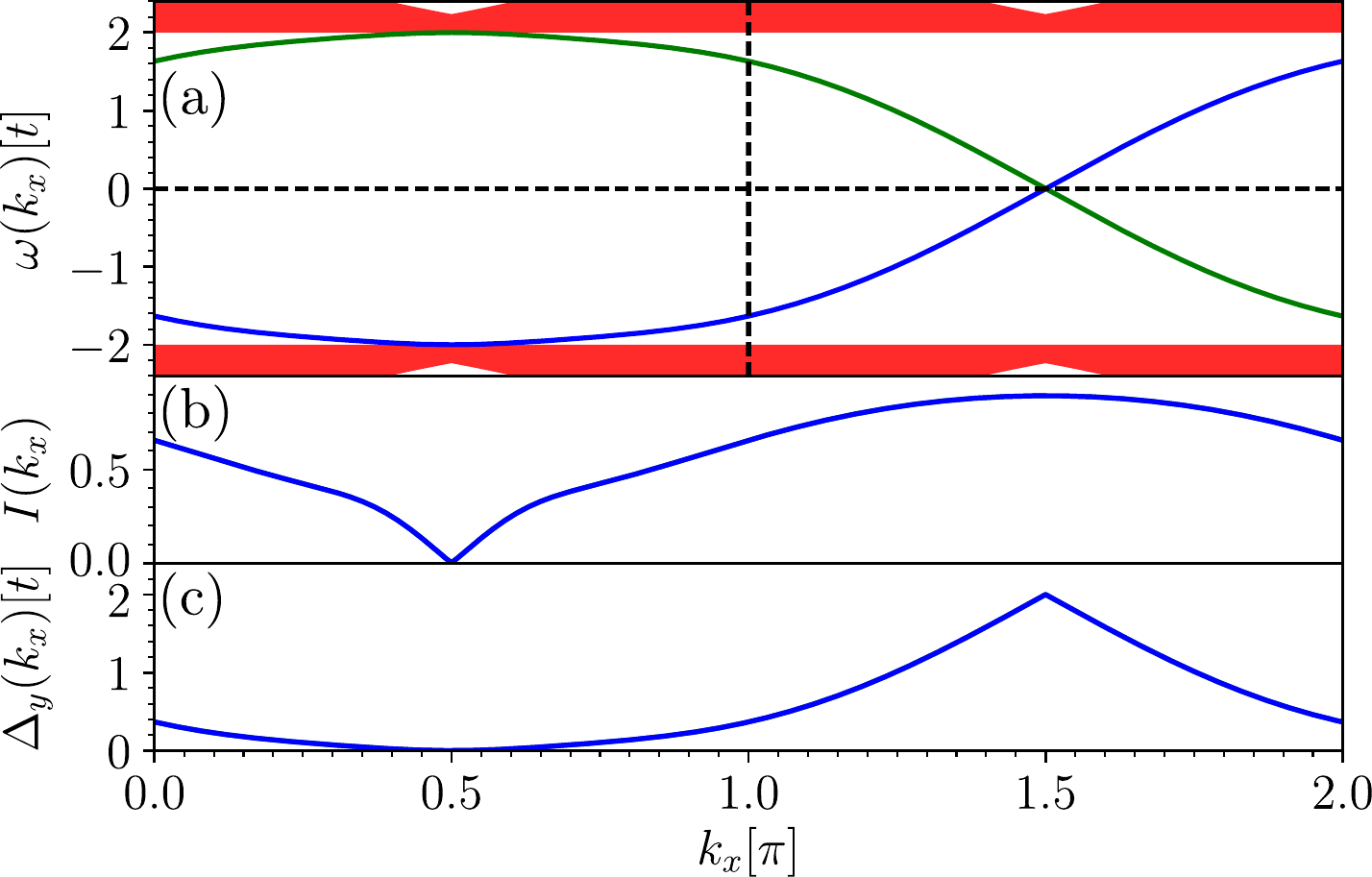}
	\caption{(a) Continua of the two bulk bands and dispersions of 
	the two in-gap states for $s = 0 t$, $d = -1 t$, and $\phi = \pi/4$. 
	The continua of the bulk bands are marked by filled red areas. 
	(b) The IPR of both edge states as function of the wave vector $k_x$. 
	(c) The indirect gap of the edge states $\Delta_y$ vs.\ the wave vector $k_x$.} 
	\label{fig:square2}
\end{figure}

For completeness, we present the localization behavior of the checkerboard model
 as function of $k_x$. As complement to the plots shown in the main text, 
Fig.\ \ref{fig:square2} displays the continua, the dispersions, the IPR,
and the indirect gap for the case where localized edge states are present
for  $s = 0 t$, $d = -1 t$, and $\phi = \pi/4$. 
The bulk continua are depicted in Fig.\ \ref{fig:square2}(a) by the red shaded areas
 while the dispersions of the right- and left-moving in-gap states are displayed in blue 
and green. The corresponding IPRs of the in-gap states are shown in Fig.\ \ref{fig:square2}(b). The IPR is finite almost over the entire Brillouin zone. This is perfectly consistent
with the finite values of the related indirect gap $\Delta_y$ in panel (c). As a result of the reflection symmetry, the blue and green curves in Fig.\ \ref{fig:sample}(b) and (c) lie on top of each other like in the original Haldane model.
 
\newpage

\begin{figure}
	\centering
		\includegraphics[width=1.0\columnwidth]{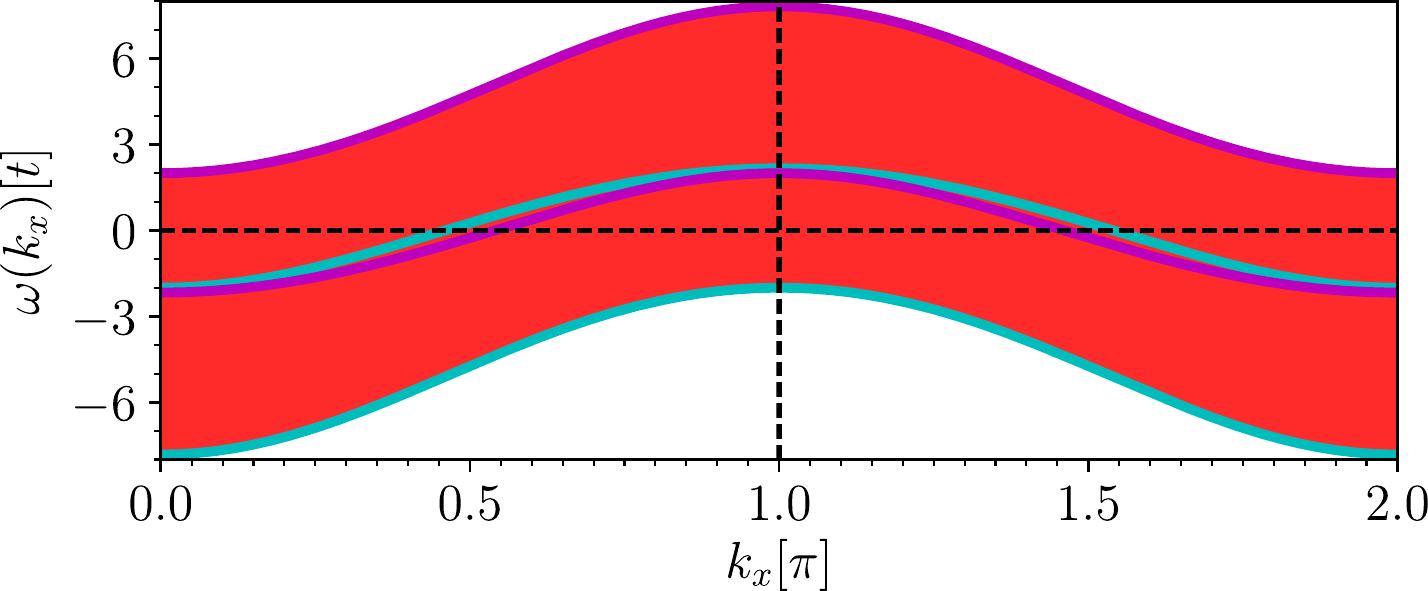}
	\caption{Continua of the two bulk bands for $s = 2.5 t$, $d = -1 t$,
	and $\phi = \pi/4$; they are marked by filled red areas. 
	The band edges of the continua are displayed in magenta for the 
	conduction band and in cyan for the valence band.}
	\label{fig:sq_crit}
\end{figure}

By tuning $s$ from $0 t$ to $2.5 t$ the indirect gap is closed as discussed in the main text.
 The continua and dispersions for $s = 2.5 t$, $d = -1 t$, and $\phi = \pi/4$ are
 plotted in \smash{Fig.\ \ref{fig:sq_crit}.} The upper and lower band overlap everywhere in the 
Brillouin zone. As a result, the indirect gap is closed and the in-gap states are 
delocalized for all wave vectors $k_x$. Hence, there are no edge states in the proper
sense of the word.

\end{document}